\documentstyle[11pt,newpasp,twoside,epsf]{article}

\def\myplotone#1#2{\centering \leavevmode
\epsfxsize=#1\textwidth \epsfbox{#2}}

\renewcommand{\textfraction}{0}

\begin{document}

\renewcommand{\textfraction}{0}
\title{A Monte Carlo Code to Investigate Stellar Collisions in Dense Galactic Nuclei.}

\author{Marc Freitag}
\affil{Observatoire de Gen{\`e}ve, CH-1290~Sauverny, Switzerland}
\author{Willy Benz}
\affil{Physikalisches Institut, Universit{\"a}t Bern, Sidlerstrasse 5,
  CH-3012~Bern, Switzerland}


\begin{abstract}
Stellar collisions have long been envisioned to be of great importance
in the center of galaxies where densities of
$10^6\,\mbox{stars}/\mbox{pc}^{3}$ or larger are attained. Not only
can they play a unique dynamical role by modifying stellar masses and
orbits, but high velocity disruptive encounters occurring in the
vicinity of a massive black hole can also be an occasional source of
fuel for the starved central engine.

In the past few years, we have been building a comprehensive table of
SPH (Smoothed Particle Hydrodynamics) collision simulations for main
sequence stars. This database is now integrated as a module into our
H{\'e}non-like Monte Carlo code. The combination of SPH collision
simulations with a Monte Carlo cluster evolution code seems ideally
suited to study the frequency, characteristics and effects of stellar
collisions during the long term evolution of galactic nuclei.
\end{abstract}

\section{Introduction}

Compact massive dark objects, with masses $10^6-10^9\,M_\odot$,
have been found in the center of nearly every bright galaxy where they
have been searched through measurements and modeling of the gas or
stellar kinematics (see reviews by {Kormendy} \& {Richstone} 1995; {Richstone} {et~al.} 1998; {Ho} 1999; {Moran}, {Greenhill}, \& {Herrnstein} 1999; {Kormendy} 2000). In the two cases with the highest resolution, i.e. the
Milky Way and NGC~4258, the size of the central object is
observationally constrained to be so small that models resorting to
compact cluster of small dark objects (Neutron stars, stellar black
holes, brown dwarfs,\ldots) seem very unlikely as such concentrations
would not survive evaporation or run-away merging for many $10^9$
years ({Maoz} 1998).  It it thus widely believed that these objects
are ``super-massive'' black holes (SBHs).

Our work is devoted to an exploration of some intriguing consequences
a SBH should have on a surrounding stellar cluster and of the
long-term evolution of such a system. Of particular interest to us are
two kinds of disruptive events which could release stellar gas in the
vicinity of the SBH and thus lead to bright accretion phases, even in
otherwise non-active nuclei. These
processes are tidal disruptions and stellar collisions. Other
mechanisms through which a stellar cluster can contribute to the
feeding of a SBH include stellar winds ({Shull} 1983; {David}, {Durisen}, \&  {Cohn} 1987b; {Norman} \& {Scoville} 1988; {Coker} \& {Melia} 1997), envelope-stripping when stars cross a
pre-existing accretion disk
(mainly relevant to red giants, see {Armitage}, {Zurek}, \& {Davies} 1996) and inspiraling
induced by strong emission of gravitational waves
(mainly relevant to compact remnants, see {Hils} \& {Bender} 1995; {Sigurdsson} \& {Rees} 1997; {Miralda-Escud{\'{e}}} \& {Gould} 2000; {Freitag} 2000). 
However, in this conference paper, we naturally focus on collisions
between main sequence (MS) stars.

To treat collisions with as much realism as possible, we decided to
determine their outcome through a comprehensive set of SPH
simulations. This important part of our work, described in
Sec.~\ref{sec:coll} ({Freitag} \& {Benz} 2000a), resulted in a database
incorporated, as a module, in a new Monte Carlo (MC) cluster evolution
code, presented in Sec.~\ref{sec:MCcode} ({Freitag} \& {Benz} 2000b). This provides
us with a numerical tool which seems ideally suited to investigate
collisions in dense stellar clusters. Although our simulations can
potentially produce detailed lists of collisionally formed objects
(such as blue-stragglers) and not only overall rates, so far we have
mainly addressed the question of the global influence of collisions on
the SBH $+$ cluster system.

Previous works relied on highly simplified prescriptions to account
for collisional effects in the stellar dynamics of galactic
nuclei\footnote{ With the noticeable exception of {Rauch} (1999) who
  used fitting formulae obtained through a limited set of collision
  simulations.}. This situation stemmed not only from the limited
knowledge of the collision itself, to be acquired from 3D
hydrodynamical simulations, but also from intrinsic limitations of the
stellar dynamics codes. Direct Fokker-Planck integrations, while very
fast, treat the stellar system as a set of continuous distribution
functions, one for each stellar mass. Thus, the mass spectrum is
discretized into a few mass classes and collision products have to be
re-distributed into these bins in a rather unphysical way. This
shortcoming is required for mergers ({Lee} 1987; {Quinlan} \& {Shapiro} 1990) or for
collisions leading to partial mass loss ({David}, {Durisen}, \&  {Cohn} 1987a; {David} {et~al.} 1987b; {Murphy}, {Cohn}, \& {Durisen} 1991);
only if complete disruption is assumed ({McMillan}, {Lightman}, \& {Cohn} 1981; {Duncan} \& {Shapiro} 1983), can it
be avoided. But this latter assumption is, by itself, a gross
over-simplification. On the other hand, $N$-body simulations can in
principle incorporate realistic collisions, but as their results can
not safely be scaled to larger $N$,\footnote{This is due to the fact
  that various processes, e.g., relaxation, evaporation, collisions,
  \dots have time scales with different dependencies on $N$.} they are
still presently restricted to systems containing a few $10^4$ stars at
most (see the work on open clusters by  {Portegies Zwart} {et~al.} 1999). Even
though the computer hardware and software dedicated to $N$-body
integration progress at high pace, this kind of simulation will still
be limited to about $10^6$ stars in a near future ({Makino} 2000).

In these proceedings, D.~De~Young presents an historical review of the
researches on stellar collisions in galactic nuclei so we need only
mention here a few issues appearing in the literature onto which we
can cast new light with our simulations. As further reading about the
role of collisions in stellar systems, we refer to {Davies} (1996),
for instance.

\begin{itemize}
  
\item Can stellar collisions amount to a significant gas source to
  fuel the central SBH?

\item Can repeated stellar mergers lead to run-away build-up of a very
  massive star, a possible precursor for a $10^2-10^3\,M_\odot$ seed
  BH? Or would this process be caught up by stellar evolution or come
  to saturation as small, relatively compact stars run across the low
  density massive star without being stopped ({Colgate} 1967)?
  
\item Is there any distinctive imprint of the collisions on the
  cluster's central density profile? Previous works predict $\rho
  \propto R^{-\alpha}$, with $\alpha \simeq 0.5$, a noticeably lower
  value than the $\alpha \simeq 1.75$ cusp expected in a
  non-collisional relaxed cluster around a SBH ({Bahcall} \& {Wolf} 1976, 1977).
  
\item Do stellar collisions produce particular stellar population in
  the center-most parts of the cluster? Can blue stragglers form
  through mergers in spite of the high relative velocities? Can
  collisions be efficient in stripping the envelopes of red giants
  (see Davies, these proceedings and  {Davies} {et~al.} 1998; {Bailey} \& {Davies} 1999)?

\end{itemize}

Our simulations still lack important features (mainly stellar
evolution and binaries) to address some of these questions but we hope
to demonstrate in Sec.~\ref{sec:simul} that they already produce
interesting results when applied to simple models and, most
importantly, that the potential of the MC code in this field is high.

\section{A Monte Carlo code for cluster dynamics}
\label{sec:MCcode}

In the past few years, we wrote a new code in order to study the
long-term ($10^9-10^{10}$\,years) evolution of galactic nuclei
consisting of $10^6-10^9$ stars. We developed a Monte Carlo scheme
based on the pioneering work of H{\'e}non  (1973). This
method, although adopted with deep modifications by Stodo{\l}kiewicz
 (1982, 1986) and now revived by Giersz
 (1998, 2000) and by Joshi and collaborators
({Joshi}, {Rasio}, \& {Portegies  Zwart} 2000; {Watters}, {Joshi}, \& {Rasio} 2000; {Joshi}, {Nave}, \& {Rasio} 1999), is not widely used. In particular, as far
as we know, ours is the first MC code designed to treat galactic
nuclei rather than globular clusters.\footnote{The MC code used by
  Shapiro and collaborators (see, e.g.,  {Duncan} \& {Shapiro} 1983) in a context
  similar to ours, was of quite different nature, somewhere in between
  H{\'e}non's method and direct Fokker-Planck integrations.} The MC
numerical scheme is nonetheless very attractive as a good compromise
between computational efficiency and physical realism (not to mention
ease of adaptation to new physical processes).

By ``efficiency'', we mean that integrating the evolution of a typical
dense central cluster with $0.5-2\times 10^6$ particles over a Hubble
time requires a few hours to a few days on a standard 400\,MHz~CPU.
This allows to carry out many simulations while varying initial
conditions and simulated physics to investigate the interplay of
various processes in such complex systems as galactic nuclei. The CPU
needed time increases with the number $N$ of particles like
$T_{\mathrm{CPU}} \propto N\log(N)$, a relation to be contrasted with
the $N^{2-3}$ scaling of ``exact'' $N$-body calculations. According to
simple extrapolations, a galactic nucleus simulation with $10^7$
particles would take only $\sim$10 CPU-days but we are presently
limited to lower $N$ by the available computer memory (160
bytes/particle).

By ``realism'', we mean that we can incorporate many important
physical processes into the simulation. Beyond 2-body relaxation which
is the core of the MC code, the ``micro-physics'' include stellar
collisions and tidal disruptions. Recently we added accretion of whole
stars induced by emission of gravitational radiation and a preliminary
treatment of stellar evolution (not covered here, see
 {Freitag} 2000). Furthermore, the MC code copes with the cluster's
self-gravitation, the growth of a central BH, an arbitrary stellar
mass spectrum and velocity distribution. As demonstrated by
{Stodo{\l}kiewicz} (1986), {Giersz} (1998) and {Rasio} (2000), the dynamical
effects of binaries can also be included in MC codes. However, in the
center of a SBH-hosting cluster, the velocity dispersion is so large
that most binaries, being ``soft'' are likely to be disrupted in
gravitational encounters with other stars, instead of acting as a heat
source like they would do in globular clusters (see, e.g.,
 {Binney} \& {Tremaine} 1987; {Spitzer} 1987).

Unfortunately, the MC scheme also suffers from a few shortcomings. The
main limitations stem from the very simplifying assumptions that make
MC codes so efficient: namely those of spherical symmetry and constant
dynamical equilibrium. Consequently, it seems very difficult if not
impossible to include such effects like BH wandering, cluster
rotation, triaxiality, interaction between stars and an accretion
disk, resonant or violent relaxation\ldots

Our code is described in detail in {Freitag} \& {Benz} (2000b). Here we just outline
its basics. The stellar cluster is represented as a set of
``particles''; each of them can be seen as a spherical shell of stars
that share the same properties, namely the stellar mass $M_{\ast}$,
orbital energy $E$, modulus of angular momentum $J$ and instantaneous
distance to the center $R$ (the radius of the shell). Together, these
particles define a smooth spherical potential $\Phi(R)$ which is
stored in a binary tree structure, for the sake of efficiency. Note
that the number of stars per particle can be set to any value but has
to be the same in each particle to ensure perfect energy conservation
in 2-body processes.

In $\Phi$, no relaxation occurs (except for a very small spurious
numerical relaxation for $N < 1000$) and the cluster is in dynamical
equilibrium. To simulate the slow relaxation-induced evolution of the
system, ``super-encounters'' (SE) are computed between particles of
adjacent ranks. A SE is a 2-body gravitational encounter between stars
from the two particles. Its deflection angle is imposed to be the RMS
value resulting, during time-step $\delta t$, from all the small angle
scatterings between stars with the properties (masses $M_1$, $M_2$
and relative velocity $V_{\mathrm{rel}}$) of the
interacting particles, i.e.,
\begin{equation}
\label{eq:thetase}
        \theta_{\mathrm{SE}} =
        \frac{\pi}{2} \sqrt{ \frac{\delta t}
        {T_{\mathrm{rel}}^{(1,2)}} }
        \mbox{\ \ with\ \ } 
        T_{\mathrm{rel}}^{(1,2)} \propto \frac{V_{\mathrm{rel}}^3}
        {\ln\Lambda G^2 n_{\ast} \left(M_1+M_2\right)^2}
\end{equation}
where $\ln\Lambda$ is the Coulomb logarithm.  A
Lagrangian radial mesh is used to evaluate the local stellar density
$n_{\ast}$.

To detect stellar collisions between two neighboring particles, we
compare a random number of uniform $[0,1[$-variate with the
probability for such an event,
\begin{equation}
\label{eq:pcoll}	
	P_{\mathrm{coll}}^{(1,2)} = \frac{\delta
	t}{T_{\mathrm{coll}}^{(1,2)}} \mbox{\ \ with\ \ }
	T_{\mathrm{coll}}^{(1,2)} = \frac{1}{
	n_{\ast}V_{\mathrm{rel}}S_{\mathrm{coll}}^{(1,2)} }.
\end{equation}
The collisional cross section reads
\begin{equation}
\label{eq:scoll}
	S_{\mathrm{coll}}^{(1,2)} = \pi
	(R_1+R_2)^2 \left[ 1 +
	\left(\frac{V^{(1,2)}_{\ast}}{V_{\mathrm{rel}}}\right)^2
	\right] \mbox{\ \ with\ \ } V^{(1,2)}_{\ast} =\left(
	\frac{2G(M_1+M_2)}{R_1+R_2} \right)^{1/2}
\end{equation} 
where $R_{1,2}$ are the stellar radii.

To increase code speed, we use $R$-variable time-steps, that are a
small fraction $\eta$ of the local relaxation and/or collision time,
$\delta t(R) \leq \eta\left( T_{\mathrm{rel}}^{-1} +
  T_{\mathrm{coll}}^{-1} \right)^{-1}$.

To check whether a particle is tidally disrupted by the central SBH or
plunges directly through the horizon, we simulate the random walk of
the tip of particle's velocity vector due to small angle scatterings
during $\delta t$. This is necessary because, as the ``loss cone''
aperture $\theta_{\mathrm{LC}}$ is tiny ({Lightman} \& {Shapiro} 1977), the time scale
for entering or leaving it, of order $\theta_{\mathrm{LC}}^2
T_{\mathrm{rel}}$, is generally much smaller than $\delta t$. Without a
procedure to ``over-sample'' the time step, we would miss a lot of
loss cone events as the velocity vector would just jump over
$\theta_{\mathrm{LC}}$.

Finally, if the particle avoided disruption, we randomly select a
position $R$ on its new orbit with a probability density that matches
the fraction of time spent at each radius, ${\mathrm{d}} P/ {\mathrm{d}} R
\propto V_{\mathrm{rad}}(R)^{-1} $. This concludes a simulation step. 
The next one starts with the random selection of another pair of
particles according to probability $\propto \delta t(R)^{-1}$.
 
\section{A comprehensive set of collision simulations between MS-stars}
\label{sec:coll}

\subsection{Approach}
\label{subsec:collappr}

In the MC scheme, the orbital and stellar properties of any given
particle are independent of those of any other particle. This means
that these quantities can be modified in any physically reasonable
way. In particular, any prescription can be used for the outcome of
stellar collisions so that we decided to describe them as
realistically as possible through results of an important set of
hydrodynamical computations of collisions between MS stars.

The numerical algorithm we use is the so-called ``Smoothed Particle
Hydrodynamics'' method (SPH, for a description see  {Benz} 1990).
As a genuinely 3D Lagrangian scheme that allows large density
contrasts and imposes no spatial symmetries or limits, it is the
method of choice to tackle this problem. This explains why the vast
majority of previous investigations in this domain were done with SPH
({Benz} \& {Hills} 1987, 1992; {Lai}, {Rasio}, \& {Shapiro} 1993; {Lombardi}, {Rasio}, \& {Shapiro} 1996, amongst others), with the
noticeable exception of the early work of {Seidl} \& {Cameron} (1972) who used a 2D
finite difference algorithm.

Our goal was to sample a region in the space of collisions' initial
conditions large enough so that most collisions happening in the
course of the simulation of a galactic nuclei would be comprised in
that domain. A reasonably good description of a collisions between MS
stars imposes a four dimensional parameter space.

The first two quantities to be specified are the stellar masses $M_1$
and $M_2$, with $M_1\le M_2$. If MS stars with different masses had
homologous internal structures (which would require a power-law
mass-radius relation, in particular), we could scale out the absolute
mass and use $q=M_1/M_2$ as the only mass parameter. But we use
realistic stellar structure models from {Schaller} {et~al.} (1992) and
{Charbonnel} {et~al.} (1999) for $M_{\ast}=0.4-85\,M_{\odot}$ and $n=1.5$
polytropes for 0.1--0.3\,$M_{\odot}$ so that we have to specify both
absolute masses independently.

In globular clusters, the velocity dispersion is of order a few
10\,km\,s$^{-1}$ which is much lower than the escape velocity from the
surface of MS stars ($\sim$ 600--1200\,km\,s$^{-1}$) so that the
relative velocity at infinity plays virtually no role in
collisions. This is not true in galactic nuclei, where
$V_{\mathrm{rel}}$ can be nearly arbitrarily high near a SBH. For
instance, velocities of 1000--1500\,km\,s$^{-1}$ have been measured
({Genzel} {et~al.} 1997; {Ghez}, {Morris}, \& {Becklin} 1999) at the Galactic center. So, $V_{\mathrm{rel}}$ 
is the next initial parameter of importance.

Finally, we have to specify the impact parameter $b$, i.e. the
distance between the trajectories of the two stars if they were
straight lines. It is often more convenient to use $d_{\mathrm{min}}$,
the periastron separation for the corresponding 2 point-mass
hyperbolic encounter. If we can neglect tidal deformation until
contact, $d_{\mathrm{min}}/(R_1+R_2) \in [0;1]$ is necessary for physical
collision. We restrict ourselves to this domain as our resolution is
probably too low to treat tidal interactions
properly.\footnote{According to {Kim} \& {Lee} (1999), the cross-section for
  formation of a tidal binary without physical contact vanishes at
  $V_{\mathrm{rel}}/V_{\ast} \simeq 0.1$ for $n=3$ polytropic
  structures and at $V_{\mathrm{rel}}/V_{\ast} \simeq 1$ for $n=1.5$.
  Consequently, such processes could occur at a significant rate in
  galactic nuclei only between stars that both have $M_{\ast} <
  0.7\,M_{\odot}$.  Furthermore, the main uncertainties in our
  understanding of a tidal binary don't reside in the conditions for
  its formation at first periastron passage (this can be delineated
  using linear oscillation theory) but in its longer term evolution
  and fate. The main issues are about the impact of the deposition of
  tidal energy on the stellar structure and the interplay between the
  stellar oscillations and the orbit of the binary. The fraction of
  tidal binaries that will quickly merge is still unknown.}

The 4D initial parameter space is thus
$(M_1,M_2,V_{\mathrm{rel}},d_{\mathrm{min}})$. Other quantities that
could affect the collisions' outcomes, like stellar rotation,
metallicity, age on MS and so on are neglected as they are probably of
second order importance. A related question of interest in highly
collisional systems where run-away merging could occur is how
collisions themselves affect the structure of stars and how these
modifications could affect further collisions. We leave any study of
this somewhat far-fetched issue, considering that it is more useful to
first assess the physical conditions required for such collisional
run-away to set in. In our cluster simulations, we assume that, after
a collision, a star immediately returns to a ``standard'' MS
structure. In fact, it takes a Kelvin-Helmholtz time scale
($T_{\mathrm{KH}}\simeq 1.6\times 10^7$\,yrs for the sun) for thermal
equilibrium to be recovered, but, in most environments
$T_{\mathrm{coll}} \gg T_{\mathrm{KH}}$, so the short
post-collision swollen phase can be neglected.

\begin{figure}
\myplotone{0.7}{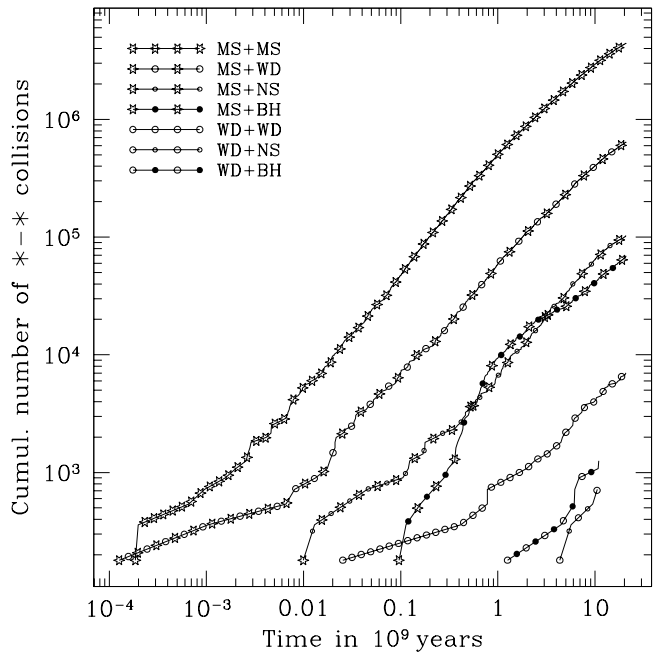}
\caption{
Cumulative number of collisions between stars of various types in a
simulation of a dense galactic nucleus model with a realistic evolved
stellar spectrum. In this MC simulation, only MS+MS collisions are
treated realistically \protect ({Freitag} 2000). Red giants are not
included.} 
\label{fig:cumncoll}
\end{figure}

Although we chose to consider only collisions between MS stars, they
may not dominate the total collision rate in many astrophysical
environments.  Indeed, as can be seen from Eq.~\ref{eq:scoll}, when
$V_{\mathrm{rel}} > V_{\ast}$, as it would occur close to a SBH, the
cross section scales like $R_{\ast}^2$ so that red giants (RG) could
participate in most collisions in spite of their low relative number
({Davies} 1996). An extension of the present work, taking into
account MS-RG collisions, is thus desirable to complement
the simulations by {Bailey} \& {Davies} (1999).  Of course, compact remnants can also
collide with MS stars or even with other compact stars. In dense
nuclei, such collisions should occur at low but non-vanishing rates,
as Fig.~\ref{fig:cumncoll} testifies. They are particularly
interesting as channels to form ``exotic'' objects.

The initial parameter space to be explored being so huge, we had to
limit the number of SPH particles per star to a relatively low value
(1000--15000) to save computer time. However, we used initial
structures with low mass particles in the stellar envelope and more
and more massive ones toward the center in order to get a satisfactory
resolution of the outer parts of the stars where the action takes place in
most collisions. Thus fractional mass loss rates as low as $10^{-4}$
can reliably be predicted. More than $14000$ collision simulations
have been computed on a local network of workstations. Such a high number
could only be attained thanks to an automatic software package we
developed to run simulation jobs on idle computers and analyze their
outcome with nearly no human intervention needed.

The result of a collision is described through a small set of
quantities: the fractional mass loss
$(M_1+M_2-{M_1}^{\prime}-{M_2}^{\prime})/(M_1+M_2)$, the new mass
ratio, the fractional loss of orbital energy and the angle of
deviation of $\vec{V}_{\mathrm{rel}}$. Note that these values
completely describe the kinematical outcome of a collision only if the
center-of-mass reference frame for the resulting star(s) (not
including ejected gas) is the same as before the collision.
Asymmetrical mass ejection violates this simplifying assumption by
giving the stars a global kick but we neglect this, in
order to reduce the complexity of the situation.

We have kept the final SPH particle configuration for (nearly) all our
simulations. This would allow us to re-analyze these files and extract
other quantities of interest, like the induced rotation, a possible
tell-tale sign of past collisions ({Alexander} \& {Kumar} 2000). Another interesting
issue is the resulting internal stellar structure. This is key to a
prediction of the subsequent evolution and observational detectability
of collision products  (see Sills, these proceedings and
 {Sills} {et~al.} 1997, 2000). Unfortunately, according to {Lombardi} {et~al.} (1999), low
resolution and use of particles of unequal masses can lead to
important spurious particle diffusion in SPH simulations so that our
models are probably not well suited for a study of the amount of
collisional mixing, for instance.

\subsection{Results}
Typical collisions will be described in {Freitag} \& {Benz} (2000a). Here we want to
give an overview of the results that may be extracted from our
database.

\begin{figure}
\plotone{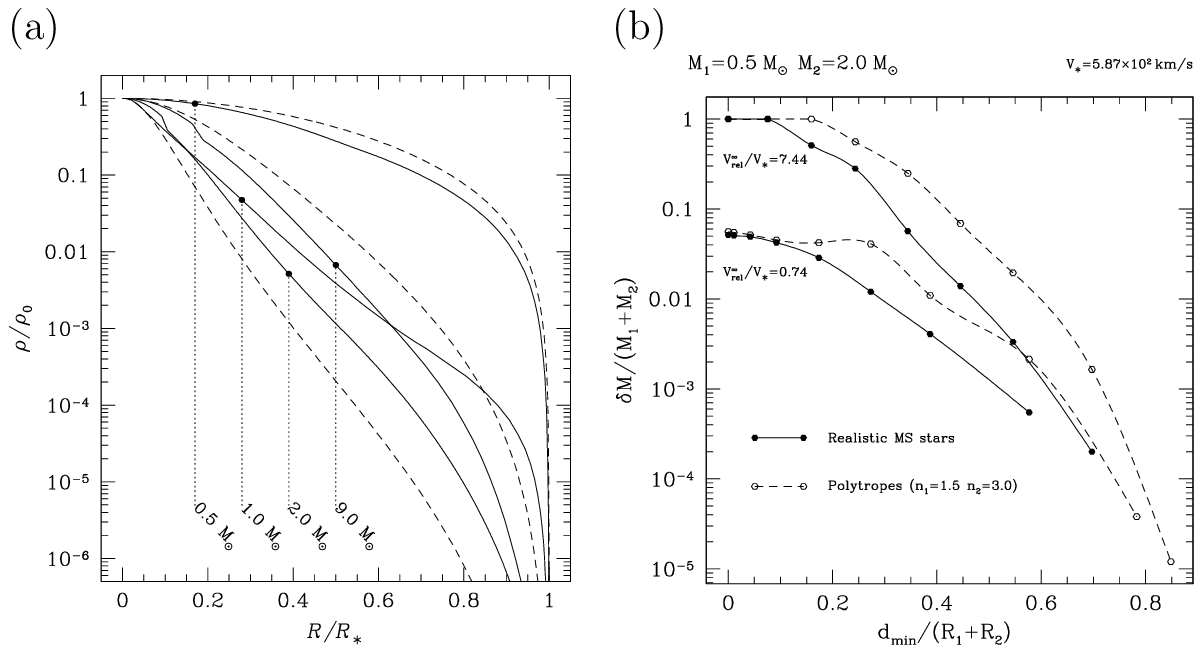}
\caption{ 
  Comparison between polytropic stars and realistic internal structure
  models. (a) Density profiles for some MS stars (solid lines) and
  $n=1.5,3,4$ polytropes (dashed lines, top to bottom). (b) Fractional
  mass loss for collisions between 0.5 and 2.0\,$M_{\odot}$ stars
  treated as polytropes or ``real'' MS stars.}
\label{fig:comppoly}
\end{figure}

\begin{figure}
\myplotone{0.94}{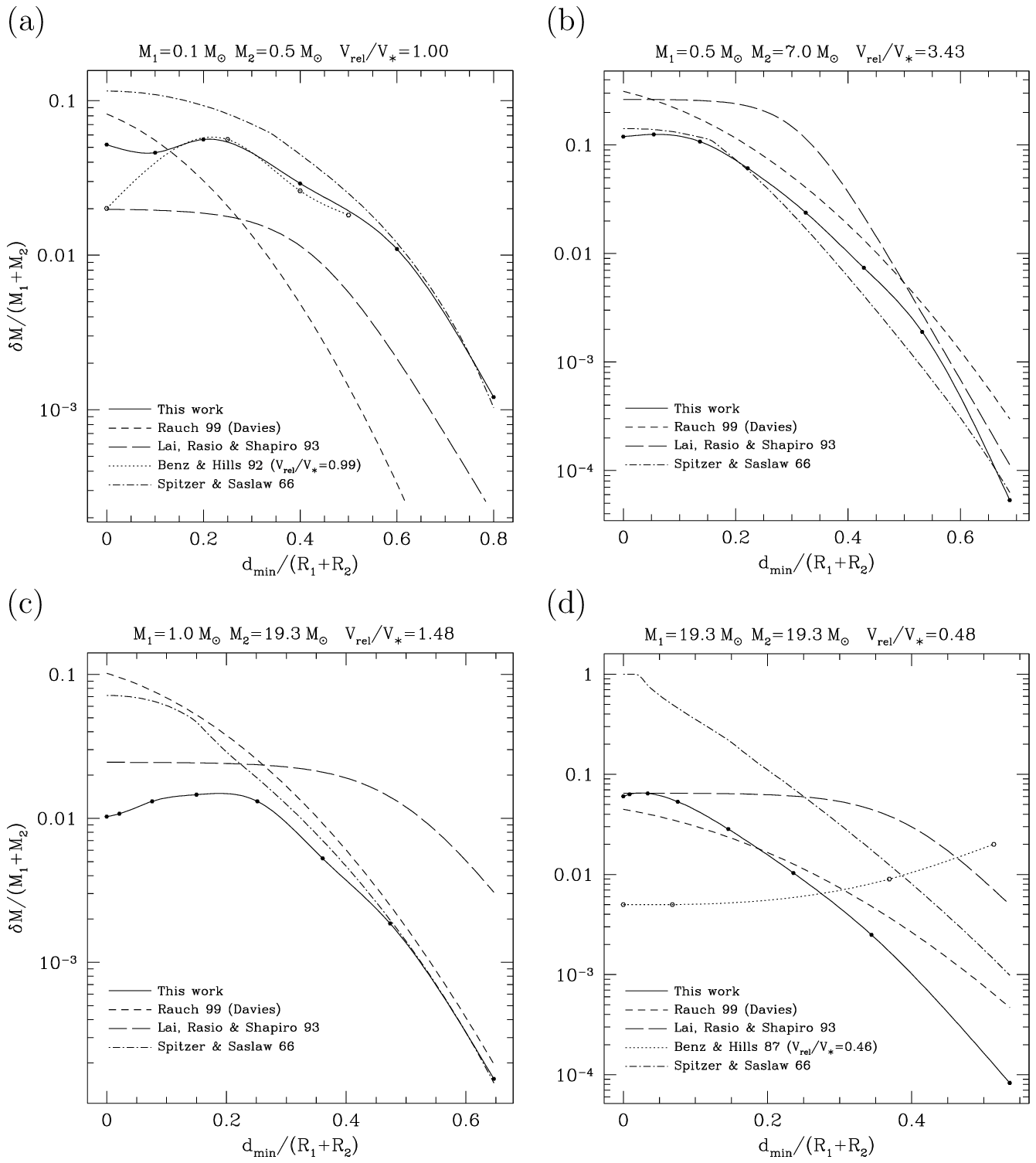}
\caption{ 
  Comparison of the fractional mass loss in some of our simulations
  with results and methods from the literature. In the case of
  \protect {Lai} {et~al.} (1993) and \protect {Rauch} (1999), we compare to
  predictions from their fitting formulae. \protect {Benz} \& {Hills} (1987, 1992)
  do not publish such formulae so we compare directly to simulation
  results. To obtain the lines labeled ``\protect {Spitzer} \& {Saslaw} (1966)'' we
  applied the semi-analytical method invented by these authors to our
  stellar models. \newline These diagrams
  clearly illustrate that sensible comparisons can only be made with
  simulation results obtained with very similar initial
  conditions and stellar structures. For instance, our results are in
  good agreement with those of Benz \& Hills for low mass stars (panel
  (a)) but completly at odds for $19.3\,M_{\odot}$ MS stars that are
  much more concentrated than their $n=1.5$ polytropes (panel (d)).
  Another example is panel~(c) where we blindly push the published
  fitting formulae to a mass ratio much lower to the values for which
  they have been devised.}
\label{fig:complit}
\end{figure}

First, looking at Fig.~\ref{fig:comppoly}, we note that using
realistic stellar structure instead of the traditional polytropic
stars has quite an important effect on the outcome of collisions. This
is mainly due to the fact that massive stars are more concentrated
than $n=3$ polytropes. Next, we ask whether computing such a high
number of simulations was worth the trouble by confronting our results
to those of the literature. In particular, we want to know how they
compare to fitting formulae devised by {Lai} {et~al.} (1993) and Davies
(used by {Rauch} 1999) to describe the results of similar but
limited sets of SPH simulations. Considering diagrams like those of
Fig.~\ref{fig:complit}, we can draw the following conclusions:
\begin{itemize}
\item Surprisingly, the simple semi-analytical prescription from
  {Spitzer} \& {Saslaw} (1966) usually gives quite accurate results for the fractional
  mass loss in the regime with $V_{\mathrm{rel}}/V_{\ast}\geq 1$ and
  $d_{\mathrm{min}}/(R_1+R_2)>0.4$.
\item As could be foreseen, empirical fitting formula must {\bf never}
  bee used to extrapolate to initial conditions outside the
  (restricted) range they originate from.
\item In particular, the stellar structure has a central role in
  determining $\delta M$. This appears clearly in
  (dis-)agreement between our results and those of {Benz} \& {Hills} (1987, 1992)
  in Fig.~\ref{fig:complit}.
\end{itemize}
Another way to state the second point is that only a mathematical
description grounded on well understood physical arguments has a
chance to have any sound predictive power when applied to a wider set
of collisions than those it derives from. For instance, it could be
that a parameterization of the ``closeness'' of the interaction that
accounts for the mass distribution inside the stars (contrary to
$d_{\mathrm{min}}/(R_1+R_2)$) could result in a good agreement between
simulations done with various stellar structures. Unfortunately, due
to the complexity of the physical processes at play during collisions,
such a ``unifying'' description seems very difficult to find and we
have failed to figure it out so far. Consequently, we tried to cover
as completely as possible the relevant domain of initial conditions
and we use an interpolation algorithm to determine the outcome of any
given collision that happens in a cluster simulation run.

\begin{figure}
\plotone{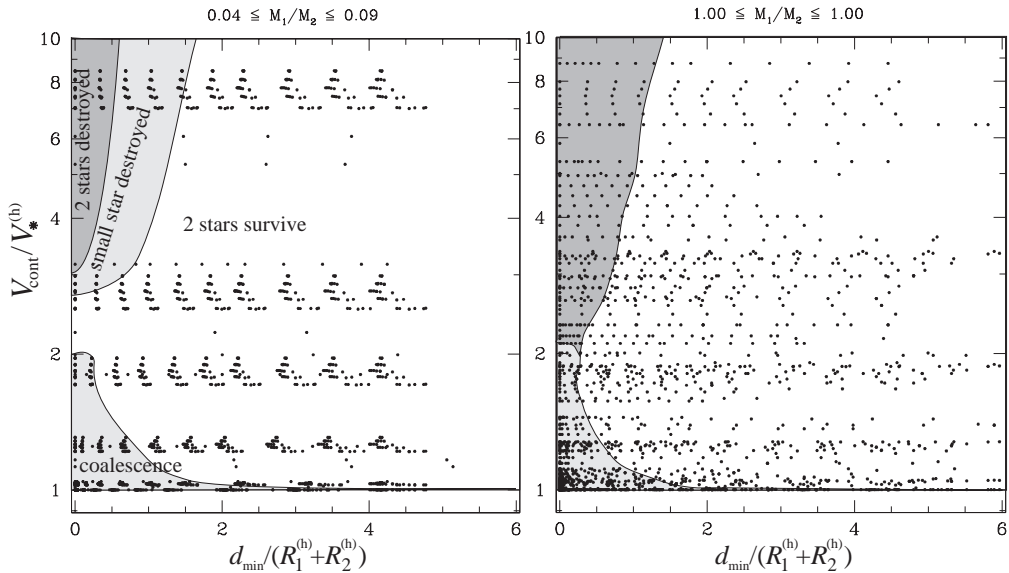}
\caption{ Diagrams for the number of surviving stars in a selection of 
  simulations. The left panel shows all simulations with $0.04 \leq
  M_1/M_2 \leq 0.09$ and the right panel all equal-mass encounters.
  Each dot stands for one SPH simulation. $R_1^{\mathrm{(h)}}$ and $R_2^{\mathrm{(h)}}$ are the
  radii that contain half the stellar mass of each star.
  $V_{\mathrm{cont}}$ is the relative velocity at ``half-mass
  contact'' (separation equal to $R_1^{\mathrm{(h)}}+R_2^{\mathrm{(h)}}$) and
  $V_{\ast}^{\mathrm{(h)}}=\sqrt{2G(M_1+M_2)/(R_1^{\mathrm{(h)}}+R_2^{\mathrm{(h)}})}$. For almost all collisions
  in the white regions, there are two outgoing stars unbound to each
  other. Collisions in light gray regions results in a merger, a bound
  binary (prone to subsequent merging) or the disruption of one star.
  Finally, dark grey shading indicates complete destruction of both
  stars.}
\label{fig:netoiles}
\end{figure}

A zero-order description of the outcome of a stellar collision
consists in the number of surviving stars. This is what we show in
Fig.~\ref{fig:netoiles}. Collisions with similar values of $q=M_1/M_2$
have been grouped on the same plot, regardless of the absolute values
of the masses.  Interestingly, in a initial conditions plane
parameterized by ``half-mass'' quantities (see caption of figure),
well defined borders appear that separate various outcome regimes. It
also appears that, unless $V_{\mathrm{rel}}$ is very low, collisions
that lead to coalescence (at low relative velocities) or complete
disruptions (at high $V_{\mathrm{rel}}$) must be nearly head-on. By
far the most likely outcome at velocities in excess of
100\,km\,s$^{-1}$ is the preservation of both stars with only a small
amount of mass loss. For small $M_1/M_2$ ratio, even head-on
collisions do not necessarily result in mergers; the small star can
fly through the large one without being stopped or destroyed. Such an
effect, due to the high-mass star being of lower density than its
small impacter was already proposed by {Colgate} (1967) to predict an
upper mass limit to the process of run-away mergings. Whether this
limiting mechanism really operates in dense stellar clusters has
to be tested in dynamical simulations. It can be suppressed by mass
segregation effects that drive most massive stars toward the center so
that most important collisions take place between two high-mass stars.

\begin{figure}
\plotone{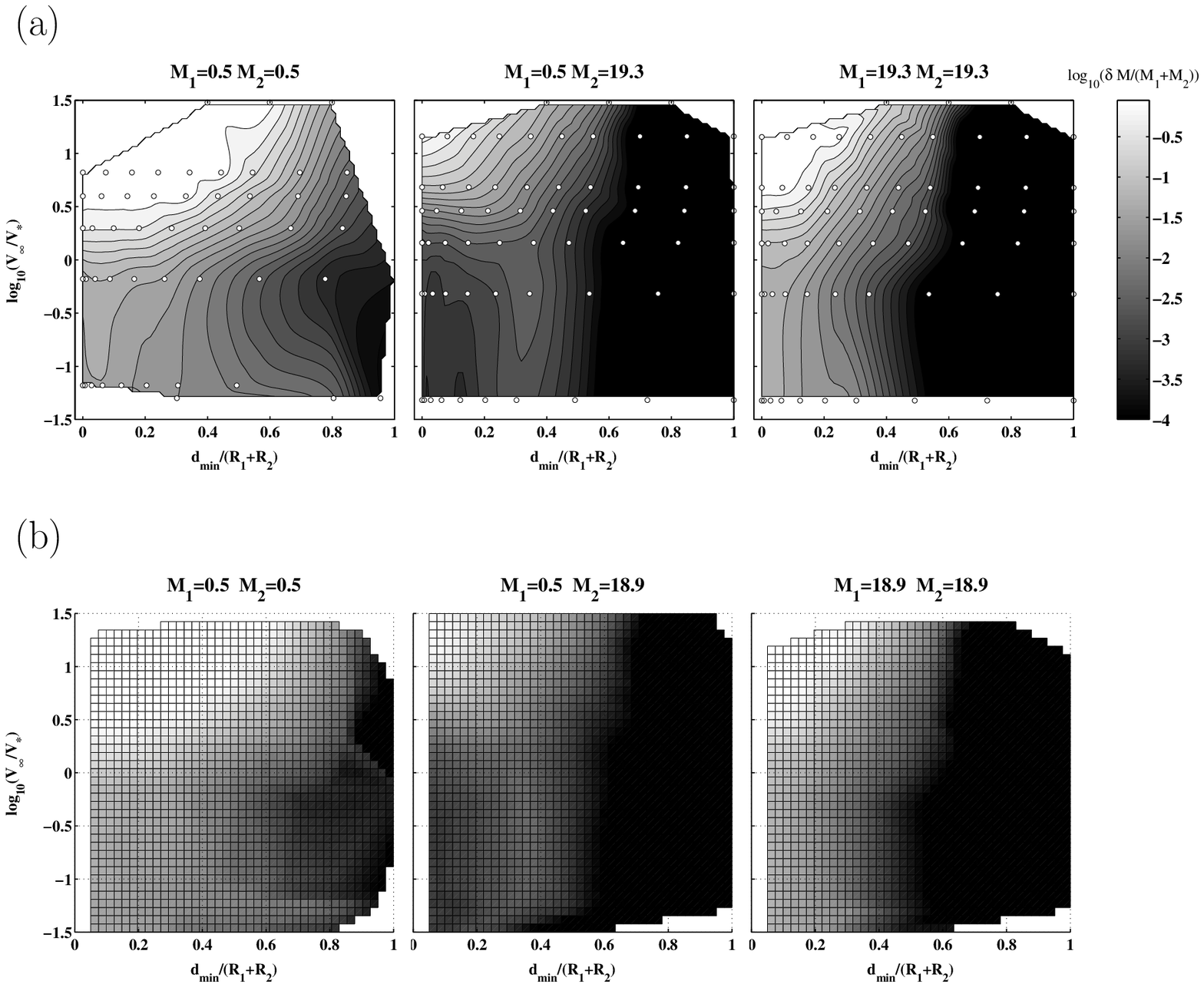}
\caption{ Collisional fractional mass loss for three different
  $(M_1,M_2)$ pairs (values in $M_{\odot}$). 
  (a) Simulation Data.
  White dots show the SPH simulations. Contours are a bicubic
  interpolation of the SPH results. 
  (b) Interpolated grid. 
  }
\label{fig:deltam}
\end{figure}

A more quantitative view on the results of 3 sets of collision
simulations is given in panel (a) of Fig.~\ref{fig:deltam} where we
show the (interpolated) fractional mass loss in the
$(d_{\mathrm{min}},V_{\mathrm{rel}})$ plane. Note how the
``landscape'' changes from one choice of $(M_1,M_2)$ to another one.
This is another indication of the difficulty of finding a universal
set of fitting formulae. The upper left white region of each diagram
indicates $\delta M/M > 85$\,\%. The small surface of this zone (in
particular for unequal masses) means that such highly destructive
events are unlikely. This is to be compared to the extent of the black
regions for which the fractional mass loss is less than $10^{-4}$.

\subsection{Integration of collisions into the MC code}

Being unable to distillate the results of our SPH simulation into any
compact mathematical formulation without losing most of the
information, we resorted to the following interpolation strategy.  In
the 4D initial parameter space, the simulations form a irregular grid
of points. We compute a Delaunay triangulation of this set using the
program \verb|Qhull|\footnote{Available at \texttt{
    http://www.geom.umn.edu/software/qhull/}} ({Barber}, {Dobkin}, \& {Huhdanpaa} 1996) which
allows us to interpolate the results onto a regular 4D grid. Three
slices in this grid are presented in Fig.~\ref{fig:deltam}(b). This
table is used in MC simulations to determine -- through a second
interpolation -- the outcome of collisions. Of course, extrapolation
prescriptions have to be specified for events whose initial conditions
fall outside the convex hull of the SPH simulation points. Most
commonly, this happens when a collisionally produced star with mass
outside the $0.1$--$74\,M_{\odot}$ range experience a further
collision. In such cases, we try to re-scale both masses while
preserving $M_1/M_2$ to get a ``surrogate collision'' lying in the
domain covered by the SPH simulations. If $V_{\mathrm{rel}}$ is too
low or too high, we increase or decrease it to enter the simulation
domain\footnote{All this fiddling does not violate mass or energy
  conservation as collision results are coded in a dimensionless
  fashion in the interpolation grid and are scaled back to the real
  physical masses and velocities before they are applied to the
  particles.}. In many cases (for instance merging at low
$V_{\mathrm{rel}}$ or complete destruction at high
$V_{\mathrm{rel}}$), this method gives very sensible results.
Encountyers with too high $d_{\mathrm{min}}$ are treated as purely
Keplerian hyperbolic deflections with no mass loss.

\section{Simulation of galactic nuclei evolution with stellar collisions}
\label{sec:simul}

To illustrate the capacities of our ``MC+SPH'' approach and the role
of collisions in the dynamics of dense galactic nuclei, we review
some results from stellar dynamical simulations of simple nuclei
models. The nominal model is a Plummer cluster with a scale radius of
$R_{\mathrm{Plum}}=0.3$\,pc that contains $5.09\times 10^7$ MS stars
with a Salpeter mass spectrum: $dN/dM_{\ast} \propto
M_{\ast}^{-2.35}$, $M_{\ast} \in [0.2,20]\,M_{\odot}$. In its center,
we put a seed black hole
($M_{\mathrm{BH}}=10^{-6}M_{\mathrm{clst}}$) which is allowed to
grow through accretion of stellar gas released in stellar collisions
and tidal disruptions. Accretion is assumed to be complete and
instantaneous. Stellar evolution is not simulated. The simulations
were realized with 512\,000 particles.

\begin{figure}
\plotone{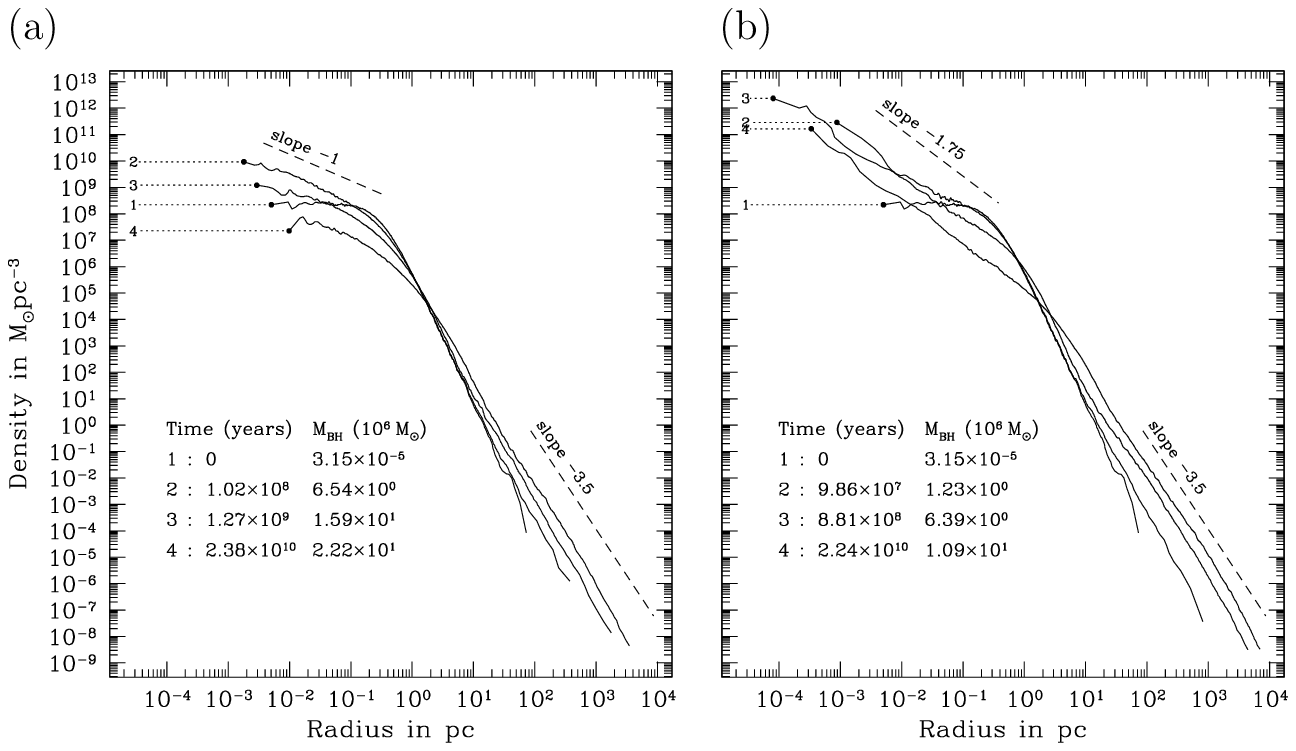}
\caption{ Evolution of the cluster's density profile for our ``nominal 
  model'' (see text). (a) Collisions assumed to be completely
  disruptive. (b) Collisions modeled through SPH generated
  interpolation grid.  }
\label{fig:compdens}
\end{figure}

\begin{figure}
\plotone{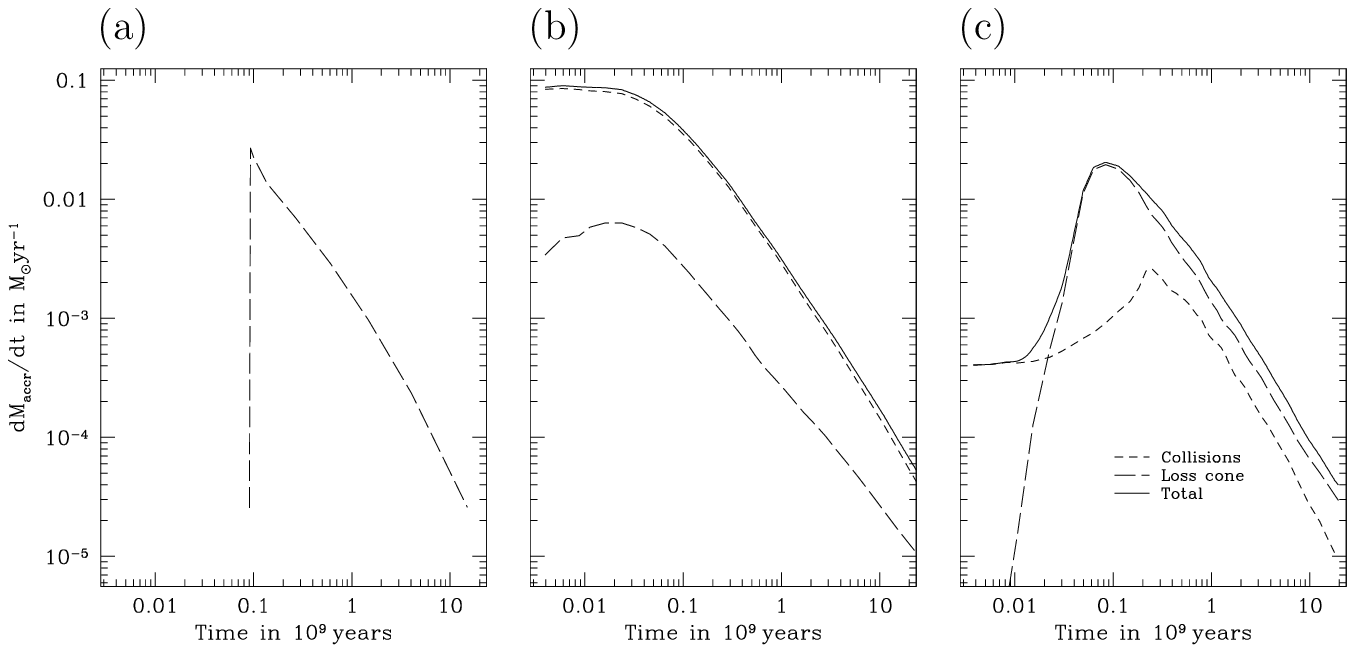}
\caption{ Accretion rate on the central BH for three variations of the 
  nominal model (see text). ``Loss Cone'' processes comprise tidal
  disruptions (dominant) and direct plunges through the BH's horizon
  (very rare). (a) No collisions. (b) Disruptive collisions. (c)
  Realistic collisions (SPH).}
\label{fig:compdmdt}
\end{figure}

Fig.~\ref{fig:compdens} shows snapshots of the density profile during
the evolution of this model. When collisions are treated
realistically, using our SPH grid, a steep central cusp with slope
$\sim -1.75$ develops. This result is very similar to what is obtained
when collisions are switched off and tidal disruptions are the only
channel to consume stars ({Bahcall} \& {Wolf} 1976, 1977). A milder slope of about
$-1$ is obtained when collisions are assumed to result in complete
stellar disruption. Even though we start with a model with very high
central density, after a Hubble time, the mass density at 0.1--1\,pc
from the BH has reached a value similar to what is measured in the
center of the Milky Way ({Genzel} {et~al.} 1997). However, at that time, the
BH's mass in our model is about $10^7\,M_{\odot}$, nearly 4 times
larger than Milky Way's value. In Fig.~\ref{fig:compdmdt}, we compare
the rates of mass accretion onto the central BH. When treated
realistically, collisions dominate over tidal disruptions only during
a short initial phase before massive stars segregate toward the
center.

\begin{figure}
\plotone{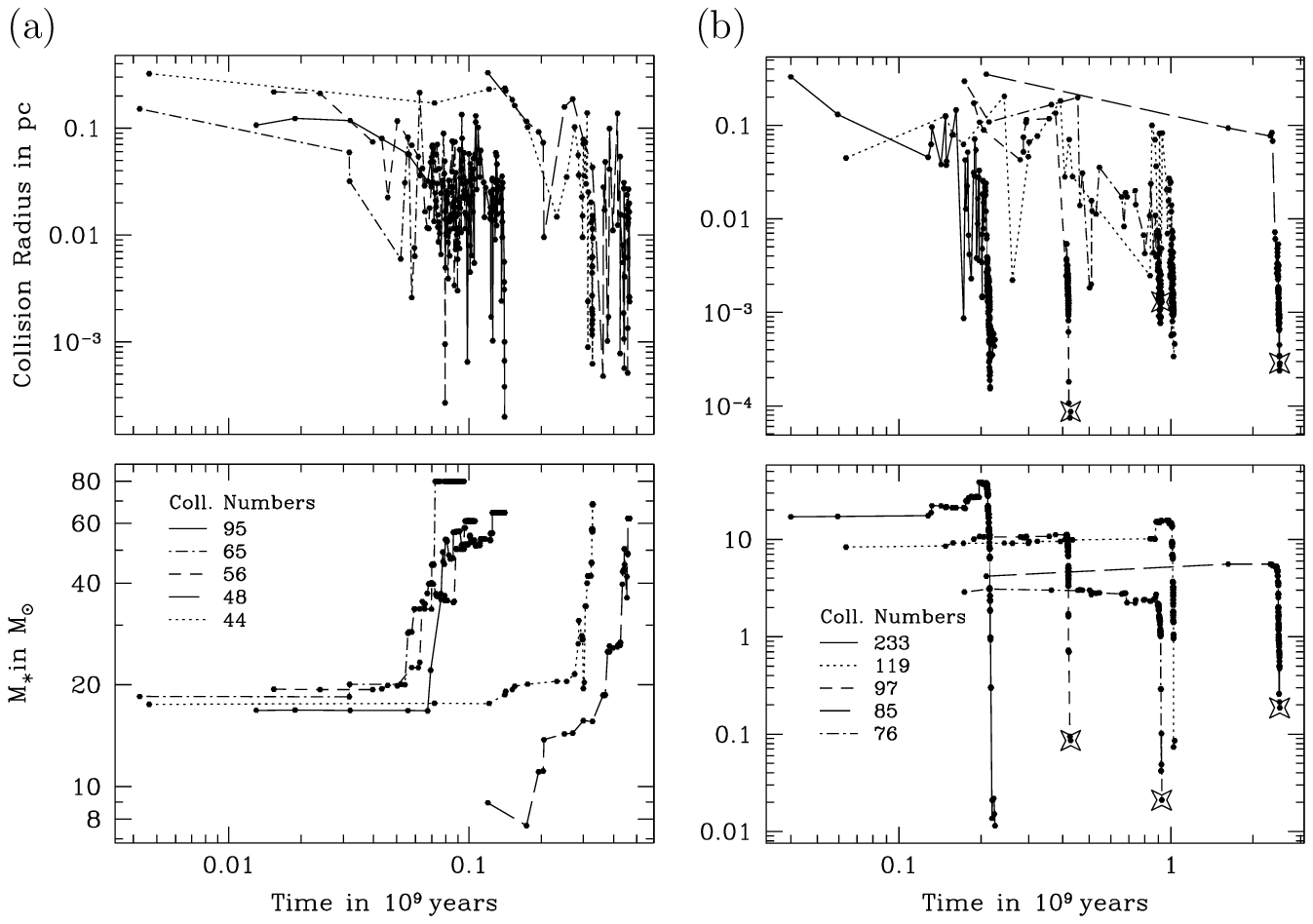}
\caption{ Collisional histories for a few stars that experienced 
  high numbers of collisions.  We show the distance to the center
  (top) and the initial stellar mass (bottom) for each successive
  collision. A final star symbol indicates the ``death'' of the
  particle, either by complete disruption or by merging with a larger
  star. (a) No central BH. (b) Central BH with initial mass
  $M_{\mathrm{BH}}=10^{-6}M_{\mathrm{clst}}$.  The labels are the
  total number of collisions for each particle.}
\label{fig:collhist}
\end{figure}

We can not only explore the structure and evolution of the stellar
cluster as a whole but also investigate some processes in more detail.
For instance, it is possible to study the properties of individual
collisions. In Fig.~\ref{fig:collhist}, we follow a selection of stars
that experienced a large number of collisions. We report the distance
to the center and the stellar mass before each collision. In a cluster
without a central BH (panel~(a)), the typical evolution of one of
these frequently colliding stars is to sink toward the center while
growing through a few mergers. In this simulation, the merging process
is not allowed if the colliding star already has a mass beyond $\sim
60\,M_{\odot}$ but there is no doubt that it would otherwise lead to
very massive stars.  Of course such results may be significantly
altered when stellar evolution is introduced. For instance, the star
represented by the dotted line would not be able to wait during $\sim
7\times 10^7$\,years between two successive mergers as the lifetime of
a $20\,M_{\odot}$ star on the MS is only of order $10^7$\,years. When
a seed BH is present initially (panel~(b)), it rapidly grows and leads
to such an important increase in the stellar velocities near the
center that mergers are totally quenched. Most collisions are then
disruptive and the average stellar mass in the central regions
actually decreases.

\section{Conclusions}

When assumptions of spherical symmetry and dynamical equilibrium are
reasonable, the Monte Carlo code for cluster dynamics appears as the
method of choice to get detailed statistical predictions about the
role and characteristics of collisions (and other physical
processes) during the evolution of a stellar system. 
The use of SPH-based prescriptions to include collisions enables us to 
take the best advantage of the flexibility of the MC scheme in terms
of realism. 

Our models still lack other important and/or interesting physical
aspects (stellar evolution, role of red giants and binaries,\ldots).
Other ingredients could be treated with more rigor. For instance, in
the same spirit of our approach of stellar collisions, we could easily
use the results of SPH simulations of tidal interactions between a
star and the SBH (e.g. {Fulbright} 1996) to determine the outcome
of these events.  For the time being, we assume them to result in
complete disruption of the star. More ``realistic'' prescriptions
would not necessarily yield more reliable results, though, as the
fraction of stripped stellar mass that is eventually accreted on the SBH
is still a matter of debate ({Ayal}, {Livio}, \& {Piran} 2000, and references therein).
This illustrates the fact that many ``improvements'' could actually
amount to adding more and more sources of uncertainty in the
simulations. In such a context, it is all the more useful to dispose
of a numerical tool flexible enough to allow changes in the treatment
of various physical effects and fast enough to allow large sets of
simulations to be conducted to test for the influence of these changes 
and the interplay between the many physical aspects of the problem.

Concerning the role of stellar collisions in the evolution of galactic
nuclei, our present results may be considered disappointing. Indeed,
even in cluster with quite extreme initial conditions (high stellar
density), collisions do not leave any strong imprint on the overall
structure of the stellar cluster. Neither do they feed the central BH
more efficiently than tidal disruption or, presumably, stellar
evolution. However, it must be stressed that collisions could have
played a role of greater importance in the past if the present day
nuclei have evolved from denser configurations. Further, more
systematic sets of simulations will allow us to delineate the
conditions leading to a collisional phase in the evolution of a
cluster. 

Furthermore, even if not efficient enough to rule the dynamics,
collisions are interesting per se. More work is required to
determinate the observational consequences of these events (creation
of ``exotic'' stars, accretion of gas onto the central BH) but our
code will stand as the central backbone for these future, more
complete, studies.

\acknowledgments

One of us (MF) wants to thank the organizers of this conference for
partial financial support to attend it. This work has been supported
in part by the Swiss National Science Foundation.


\end{document}